\documentclass[letterpaper]{article}

\usepackage{graphicx}  

\usepackage[utf8]{inputenc} 
\usepackage[T1]{fontenc}

\begin{document}

\title{Splitting the atom before the atom had been split}

\author{
{\Large Julyan H. E. Cartwright$^{1,2}$} \\
$^1$ Instituto Andaluz de Ciencias de la Tierra, \\ CSIC, 18100 Armilla, Granada, Spain \\
$^2$ Instituto Carlos I de F\'{\i}sica Te\'orica y Computacional, \\ Universidad de Granada, 18071 Granada, Spain
\\ \\ julyan.cartwright@csic.es
}

\date{The history of the discovery of nuclear fission and how science fiction had anticipated it, published November 2024 in Czech translation in \emph{Krakatit 100},  a celebration of the novel's centenary,  Jitka \v{C}ejkov\'a, editor.}

\maketitle

In the 1920s, Karel \v{C}apek wrote two novels of science fiction dealing with a future in which nuclear fission---splitting the atom---was commonplace. \emph{The Absolute at Large} was published in 1922, and  \v{C}apek set it in 1943 \cite{Capek1922}. \emph{Krakatit} was published in 1924 \cite{Capek1924}, and set in some indeterminate time that seems similar to the other. The atom was split in his lifetime, but he did not live to see either nuclear energy or nuclear weapons, the background themes of these two novels, come to fruition. Both \emph{Krakatit} and \emph{The Absolute at Large} emerged from  \v{C}apek's interest in and familiarity with contemporary science, in particular with the energy within the atom. 

In the 1924 novel \emph{Krakatit}---the name is derived from the Indonesian volcano Krakatoa that erupted explosively in 1883--- \v{C}apek takes on the possibilities for the warlike uses of atomic energy: its explosive power latent within an atom. In \emph{Krakatit} (what we would term) atomic bombs, are smaller than those the world has today. Small, so-called suitcase nuclear weapons have been developed, and much effort was devoted to trying to track down missing Soviet ones after the fall of the Soviet Union. They are probably not common owing to these risks.

In \emph{The Absolute at Large} from 1922, on the other hand, he considers the peaceable uses of atomic energy for power generation. In the novel, a side effect of atomic transmutation is the release of ``the absolute'', a literal ``god particle'' that, by upending natural law and working miracles, wreaks havoc on humanity. Today we have a subatomic particle, the Higgs boson, that has been given that epithet in a metaphorical use of the term. However, the absolute is more like the pre-Socratic scientist Anaxagoras' ``nous'' from 2500 years earlier, which Anaxagoras conceived as some sort of god-like initial condition for the universe \cite{Cartwright2024}. 

His nuclear power plants, or atomic piles, as they used to be called, are, like his bombs, smaller than commonplace today. Nuclear microreactors do exist. Although in the 1950s there were predictions of these being used in all sorts of places, like in  \emph{The Absolute at Large}, that that never took place is probably owing to governmental regulations on nuclear devices.

 \v{C}apek was not alone among writers in looking at contemporary science and predicting the coming of atomic energy and weapons. In the decade prior, H. G. Wells, in his 1914 novel \emph{The World Set Free} \cite{Wells1914}, had already felt able to speculate on a future involving energy from the atom: ``The problem which was already being mooted by such scientific men as Ramsay, Rutherford, and Soddy, in the very beginning of the twentieth century, the problem of inducing radio-activity in the heavier elements and so tapping the internal energy of atoms, was solved by a wonderful combination of induction, intuition, and luck by Holsten so soon as the year 1933'', he wrote. ``Given that knowledge'', he said, ``mark what we should be able to do! We should not only be able to use this uranium and thorium; not only should we have a source of power so potent that a man might carry in his hand the energy to light a city for a year, fight a fleet of battleships, or drive one of our giant liners across the Atlantic; but we should also have a clue that would enable us at last to quicken the process of disintegration in all the other elements, where decay is still so slow as to escape our finest measurements. Every scrap of solid matter in the world would become an available reservoir of concentrated force.'' But also war ``Certainly it seems now that nothing could have been more obvious to the people of the early twentieth century than the rapidity with which war was becoming impossible. And as certainly they did not see it. They did not see it until the atomic bombs burst in their fumbling hands''.

 The American physicist Robert W. Wood published \emph{The Man Who Rocked the Earth} with Arthur Train in 1915 \cite{Train&Wood1915}, another novel describing the detonation of an atomic bomb, ``A single ounce of uranium contains about the same amount of energy that could be produced by the combustion of ten tons of coal---but it won't let the energy go. Instead it holds on to it, and the energy leaks slowly, almost imperceptibly, away, like water from a big reservoir tapped only by a tiny pipe. `Atomic energy' Rutherford calls it. Every element, every substance, has its ready to be touched off and put to use. The chap who can find out how to release that energy all at once will revolutionize the civilized world. It will be like the discovery that water could be turned into steam and made to work for us---multiplied a million times. If, instead of that energy just oozing away and the uranium disintegrating infinitesimally each year, it could be exploded at a given moment you could drive an ocean liner with a handful of it. You could make the old globe stagger round and turn upside down! Mankind could just lay off and take a holiday.''

When was the atom in fact split? That is not a completely straightforward question. It begins with photography. In 1857, Abel Ni\'epce de Saint-Victor was working on developing colour photography using light-sensitive metal salts, including uranium salts. By 1861, he had found that uranium salts were producing ``un rayonnement invisible \`a nos yeux''; that is ``a radiation that is invisible to our eyes'' \cite{Saint-Victor1861}. In 1868, Edmond Becquerel published \emph{La lumi\`ere: ses causes et ses effets} (Light: its causes and its effects) \cite{Becquerel1861}, a book in which he mentioned Saint-Victor's findings; specifically, that uranium nitrate could expose photographic plates without any light present. There things remained, until his son, Henri Becquerel, began in 1896 to experiment with uranium salts, initially with the idea that they might emit radiation when stimulated by sunlight. The younger Becquerel soon found that external radiation was not necessary and that the uranium salt itself was emitting something \cite{Becquerel1896}. 

This work was being carried on in Paris in the excitement after Wilhelm R\"ontgen's 1895 discovery of X rays \cite{Martins2021}. At the same time, 1896, Silvanus Thompson was looking into the same phenomenon in London: ``the persistent emission by certain substances, notably by metallic uranium and its salts, of invisible rays which closely resemble R\"ontgen rays in their photographic action, and in their power of penetrating aluminium'' \cite{Thompson1896}. Thompson wrote to G. G. Stokes about his finding, which he called hyper-phosphorescence, and Stokes had to give him the bad news  ``I fear that you have already been anticipated. See Becquerel, Comptes Rendus for Feb. 24, p. 420, and some papers in two or three meetings preceding that'' \cite{Larmor1907}.

Marie Curie, also in Paris, began to study this peculiar behaviour of uranium compounds in 1896, and soon her husband Pierre joined her research into the uranium minerals pitchblende and torbernite, which she found displayed the phenomenon to a greater extent with uranium itself.  ``The fact is very remarkable, and leads to the belief that these minerals may contain an element which is much more active than uranium'', she wrote \cite{Skwarzec2011}. The term radioactivity was coined by them as French radio-actif. ``I will call radioactives the substances that emit Becquerel rays. The name hyperphosphorescence that had been proposed for the phenomenon seems to me to convey a wrong idea about its nature'' \cite{Curie1899}. Henri Becquerel and Marie and Pierre Curie won the Nobel Prize in physics in 1903 ``in recognition of the extraordinary services he has rendered by his discovery of spontaneous radioactivity'', and ``in recognition of the extraordinary services they have rendered by their joint researches on the radiation phenomena discovered by Professor Henri Becquerel''.

Ernest Rutherford and his student Frederick Soddy at McGill University in Montreal were the first to understand that radioactivity resulted in what the alchemists had sought in centuries prior, the transmutation of one element to another.  ``An enormous store of latent energy is resident in the atoms of the radio-elements'', wrote Rutherford \cite{Rutherford1904}. They realized that there are three types of radiation involved, which they called alpha, beta, and gamma \cite{Rutherford1904,Soddy1904}. Soddy together with William Ramsay at University College London showed that as radium decayed it produced the inert gas helium \cite{Ramsay&Soddy1903}. Ramsay received the Nobel Prize in Chemistry in 1904 ``in recognition of his services in the discovery of the inert gaseous elements in air''. Rutherford was awarded the 1908 Nobel Prize in Chemistry ``for his investigations into the disintegration of the elements, and the chemistry of radioactive substances'' (he commented ``I am very startled at my metamorphosis into a chemist''). Soddy received the 1921 Nobel Prize in Chemistry ``for his contributions to our knowledge of the chemistry of radioactive substances, and his investigations into the origin and nature of isotopes''.

In 1919 Rutherford, by this time working in Manchester, created an artificial nuclear reaction bombarding nitrogen gas with alpha particles. Some nitrogen was transmuted into oxygen, with hydrogen nuclei being emitted \cite{Rutherford1919}. In 1920 he proposed the name proton \cite{Romer1997}. By 1932, Rutherford was now in Cambridge.  John Cockcroft and Ernest Walton bombarded lithium with protons in his lab. The transmutation produced two alpha particles---helium nuclei \cite{Cockcroft&Walton1932}.  At almost the same time in Rutherford's Cambridge laboratory, James Chadwick ran experiments on hitting beryllium with alpha particles \cite{Chadwick1932a,Chadwick1932b}. This emitted another particle, the neutron. The Nobel Prize in Physics 1935 was awarded to Chadwick ``for the discovery of the neutron'', and Cockcroft and Walton won the 1951 Nobel Prize in Physics ``for their pioneer work on the transmutation of atomic nuclei by artificially accelerated atomic particles''.

However Rutherford, quoted in \emph{The Times} newspaper, 12 September 1933, was dismissive of the practical applications: ``We might in these processes obtain very much more energy than the proton supplied, but on the average we could not expect to obtain energy in this way. It was a very poor and inefficient way of producing energy, and anyone who looked for a source of power in the transformation of the atoms was talking moonshine'' \cite{Rutherford1933}.

But not all scientists were of the same opinion. The day after Rutherford's pronouncement, Leo Szilard was waiting to cross the road in London: ``...as I was waiting for the light to change and as the light changed to green and I crossed the street, it suddenly occurred to me that if we could find an element which is split by neutrons and which would emit two neutrons when it absorbed one neutron, such an element, if assembled in sufficiently large mass, could sustain a nuclear chain reaction''. Szilard recounted that he had recently read \emph{The World Set Free}: ``Knowing what this [chain reaction] would mean---and I knew it because I had read H. G. Wells---I did not want this patent to become public'' \cite{Weart&Szilard1978}. His 1934 patent described how neutrons might be used to create a nuclear chain reaction, and that a critical mass of such a substance would explode. 

1934 saw Ir\`ene Joliot-Curie and Fr\'ed\'eric Joliot obtain radioactive elements by transmuting others---radioactive nitrogen from boron, radioactive phosphorus from aluminium, and radioactive silicon from magnesium---in the first demonstration of artificial radioactivity from isotopes unknown in nature  \cite{Joliot1934a,Joliot1934b}. The Nobel Prize in Chemistry 1935 was awarded to Joliot and Joliot-Curie ``in recognition of their synthesis of new radioactive elements''.

That same year, 1934, in Rome, Enrico Fermi bombarded elements in the higher reaches of the periodic table with neutrons \cite{Fermi1934a,Fermi1934b}. The 1938 Nobel prize in physics was awarded to him ``for his demonstrations of the existence of new radioactive elements produced by neutron irradiation, and for his related discovery of nuclear reactions brought about by slow neutrons''.

Ida Noddack, working in Germany, published a paper in 1934 \cite{Noddack1934} that argued of Fermi's experiments ``it is conceivable that the nucleus breaks up into several large fragments, which would of course be isotopes of known elements but would not be neighbors of the irradiated element.'' 

Also in Germany, Otto Hahn, Lise Meitner, and Fritz Strassman repeated and extended Fermi's experiments over 1934--1938. ``Lise Meitner and I decided to repeat Fermi's experiments'', he wrote \cite{Hahn1946}. Meitner, in a 1939 letter to Nature with her nephew Otto Frisch described the process of nuclear fission \cite{Meitner&Frisch1939}.  In 1945 Hahn was awarded the 1944 Nobel Prize in Chemistry---notoriously without Meitner \cite{Crawford1997}---``for his discovery of the fission of heavy atomic nuclei''.

 \v{C}apek died in 1938, just before the Second World War. In 1939, Albert Einstein and Szilard wrote a letter to US President Roosevelt on the possibility of exploiting a nuclear chain reaction to make an atomic bomb \cite{Einstein&Szilard1939}. With Fermi and Szilard, the first atomic reactor, the Chicago Pile, achieved criticality in December 1942. The first atomic bomb exploded over the New Mexico desert in July 1945.

\bibliographystyle{unsrt}
\bibliography{krakatit} 

\begin{thebibliography}{10}

\bibitem{Capek1922}
K.~\v{C}apek.
\newblock {\em The Absolute at Large (Tov\'arna na absolutno)}.
\newblock 1922.

\bibitem{Capek1924}
K.~\v{C}apek.
\newblock {\em Krakatit}.
\newblock 1924.

\bibitem{Cartwright2024}
J.~H.~E. Cartwright.
\newblock 2500 years ago scientific theories of the origin of life arose in
  ancient {Greece}.
\newblock {\em Discover Life}, 54(1):1--6, 2024.

\bibitem{Wells1914}
H.~G. Wells.
\newblock {\em The World Set Free}.
\newblock 1914.

\bibitem{Train&Wood1915}
A.~C. Train and R.~W. Wood.
\newblock {\em The Man Who Rocked the Earth}.
\newblock 1915.

\bibitem{Saint-Victor1861}
A.~N. de~Saint-Victor.
\newblock Cinqui\`eme m\'emoire sur une nouvelle action de la lumi\`ere.
\newblock {\em Comptes rendus hebdomadaires des s\'eances de l'Acad\'emie des
  sciences}, 53:33--35, 1861.

\bibitem{Becquerel1861}
E.~Becquerel.
\newblock {\em La lumi\`ere, ses causes et ses effets}.
\newblock 1861.

\bibitem{Becquerel1896}
H.~Becquerel.
\newblock Sur les radiations invisibles \'emises par les corps phosphorescents.
\newblock {\em Comptes rendus hebdomadaires des s\'eances de l'Acad\'emie des
  sciences}, 122:501--503, 1896.

\bibitem{Martins2021}
R.~de~Andrade~Martins.
\newblock A pool of radiations: {B}ecquerel and {P}oincar\'e's conjecture.
\newblock In {\em Historical Essays on Radioactivity}. Extrema: Quamcumque
  Editum, 2021.

\bibitem{Thompson1896}
S.~P. Thompson.
\newblock On hyperphosphorescence.
\newblock {\em Report of the 66th Meeting of the British Association for the
  Advancement of Science}, 713, 1896.

\bibitem{Larmor1907}
J.~Larmor, editor.
\newblock {\em Memoir and scientific correspondence of the late Sir George
  Gabriel Stokes}.
\newblock Cambridge University Press, 1907.

\bibitem{Skwarzec2011}
B.~Skwarzec.
\newblock {Maria Sk\l{}odowska-Curie} (1867--1934)---her life and discoveries.
\newblock {\em Analytical and Bioanalytical Chemistry}, 400:1547--1554, 2011.

\bibitem{Curie1899}
M.~Curie.
\newblock Les rayons de {Becquerel} et le polonium.
\newblock {\em R\'evue G\'en\'erale des Sciences}, 10:41--50, 1899.

\bibitem{Rutherford1904}
E.~Rutherford.
\newblock Radio-activity, 1904.

\bibitem{Soddy1904}
F.~Soddy.
\newblock Radio-activity, 1904.

\bibitem{Ramsay&Soddy1903}
W.~Ramsay and F.~Soddy.
\newblock Experiments in radioactivity, and the production of helium from
  radium.
\newblock {\em Proceedings of the Royal Society}, 72:204--207, 1903.

\bibitem{Rutherford1919}
E.~Rutherford.
\newblock Collision of $\alpha$ particles with light atoms {II}. {V}elocity of
  the hydrogen atom.
\newblock {\em Philosophical Magazine series 6}, 37(222):562--571, 1919.

\bibitem{Romer1997}
A.~Romer.
\newblock Proton or prouton?: {R}utherford and the depths of the atom.
\newblock {\em American Journal of Physics}, 65(8):707--716, 1997.

\bibitem{Cockcroft&Walton1932}
J.~D. Cockcroft and E.~T.~S. Walton.
\newblock Experiments with high velocity positive ions. {II}. {The}
  disintegration of elements by high velocity protons.
\newblock {\em Proceedings of the Royal Society A}, 137:229--242, 1932.

\bibitem{Chadwick1932a}
J.~Chadwick.
\newblock Possible existence of a neutron.
\newblock {\em Nature}, 129(3252):312, 1932.

\bibitem{Chadwick1932b}
J.~Chadwick.
\newblock The existence of a neutron.
\newblock {\em Proceedings of the Royal Society A}, 136(830):692--708, 1932.

\bibitem{Rutherford1933}
E.~Rutherford.
\newblock Times, 12 {S}eptember 1933.

\bibitem{Weart&Szilard1978}
S.~R. Weart and G.~W. Szilard, editors.
\newblock {\em {Leo Szilard}: His Version of the Facts. {S}elected
  Recollections and Correspondence}.
\newblock 1978.

\bibitem{Joliot1934a}
F.~Joliot and I.~Joliot-Curie.
\newblock Un nouveau type de radioactivit\'e.
\newblock {\em Comptes rendus hebdomadaires des s\'eances de l'Acad\'emie des
  sciences}, 198:254--256, 1934.

\bibitem{Joliot1934b}
F.~Joliot and I.~Joliot-Curie.
\newblock Artificial production of a new kind of radio-element.
\newblock {\em Nature}, 133(3354):201--202, 1934.

\bibitem{Fermi1934a}
E.~Fermi.
\newblock Radioattivit\`a indotta da bombardamento di neutroni.
\newblock {\em La Ricerca Scientifica}, 1(5):283, 1934.

\bibitem{Fermi1934b}
E.~Fermi, E.~Amaldi, O.~d'Agostino, F.~Rasetti, and E.~Segre.
\newblock Artificial radioactivity produced by neutron bombardment.
\newblock {\em Proceedings of the Royal Society A}, 146(857):483, 1934.

\bibitem{Noddack1934}
I.~Noddack.
\newblock {\"U}ber das {E}lement 93.
\newblock {\em Angewandte Chemie}, 47(37):653--655, 1934.

\bibitem{Hahn1946}
O.~Hahn.
\newblock {\em From the natural transmutations of uranium to its artificial
  fission}.
\newblock Nobel Lecture, 1946.

\bibitem{Meitner&Frisch1939}
L.~Meitner and O.~R. Frisch.
\newblock Disintegration of uranium by neutrons: a new type of nuclear
  reaction.
\newblock {\em Nature}, 143(3615):239, 1939.

\bibitem{Crawford1997}
E.~Crawford, R.~L. Sime, and M.~Walker.
\newblock A {Nobel} tale of postwar injustice.
\newblock {\em Physics Today}, 50(9):26--32, 1997.

\bibitem{Einstein&Szilard1939}
A.~Einstein and L.~Szilard.
\newblock {Letter to US President F. D. Roosevelt}, 1939.

\end{thebibliography}

\end{document}